\documentstyle[12pt]{article}

\begin{document}

\author{ B. Boisseau\thanks{E-mail : boisseau@celfi.phys.univ-tours.fr},
C. Charmousis\thanks{E-mail : christos@celfi.phys.univ-tours.fr}
and B. Linet\thanks{E-mail : linet@celfi.phys.univ-tours.fr} \\
\small Laboratoire de Mod\`{e}les de Physique Math\'{e}matique \\
\small CNRS/EP 93, Universit\'{e} Fran\c{c}ois Rabelais \\
\small Facult\'{e} des Sciences et Techniques \\
\small Parc de Grandmont 37200 TOURS, France } 

\title{\bf Electrostatic self-force in a static weak gravitational field
with cylindrical symmetry}

\date{}

\maketitle

\begin{abstract}

We determine the electrostatic self-force for a point charge at
rest in an arbitrary static metric with cylindrical symmetry in the
linear approximation in the Newtonian constant. In linearised Einstein theory, 
we express it in terms of the components of the energy-momentum tensor.

\end{abstract}

\thispagestyle{empty}

\newpage
\section{Introduction}

DeWitt and DeWitt \cite{dw1} have investigated the DeWitt-Brehme equation of 
motion \cite{dw2} for a point charge in a static weak gravitational field with
spherical symmetry. They determine the electromagnetic self-force for a point
charge falling non relativistically. It is not connected with the emission of
electromagnetic radiation and it is non-zero for a point charge at rest.
However, for a point charge at rest in a static gravitational field within  the
Newtonian approximation, another approach due to Vilenkin \cite{vi1} gives 
directly the induced electrostatic self-force. An exact determination of this
in the Schwarzchild spacetime has been performed \cite{sm1,ze,le} by using
the expression in closed form of the electrostatic potential \cite{li1}.

A renewed interest about the electrostatic self-force has appeared in
the studies of the
straight cosmic strings in general relativity \cite{vi2}.
For an infinitely thin straight cosmic string, the
determination of the electrostatic self-force has been performed by 
Linet \cite{li2,li3} and Smith \cite{sm2}.
 However, some  questions occur about the boundary
conditions of the electrostatic field at the position of the singular
source of the spacetime \cite{ka}. Therefore, electrostatics in the case of 
an extended straight cosmic string have been analysed \cite{pe,al} in order to
derive an expression of the electrostatic self-force.

The purpose of the present paper is to determine 
the electrostatic self-force
for a point charge at rest in an arbitrary static metric with cylindrical
symmetry. This class of spacetimes can describe extended cosmic strings in
general relativity but also different types of cosmic strings eventually
in other gravitational metric theories. Here, we limit ourselves
to the calculation of the electrostatic self-force 
in  the linear approximation in the Newtonian constant $G$ (we set $c=1$).
It can be derived from a self-interaction potential energy $W$ ; in 
Minkowskian-like coordinates we have
\begin{equation}
f^{A}=-\frac{\partial W}{\partial x^{A}} \quad A=1,2 \quad
{\rm and} \quad f^{3}=0
\end{equation}

We emphasize that we find the electrostatic potential beyond the Newtonian
approximation in which the components of the stress tensor are
neglected before the energy density. We will see that the knowledge of
the electrostatic potential in the case of a straight cosmic string
plays a central role in our method enabling us to obtain the
result in integral form.

The plan of the paper is as follows. In Section 2, we indicate the coordinate
system in which we analyse the static metrics with cylindrical symmetry. We
show in Section 3 how the determination of the electrostatic potential
in the linear approximation in $G$ can be reduced in solving two equations. In
Sections 4 and 5, we solve successively these equations obtaining as a
consequence the self-interaction potential energy. We add some concluding
remarks in Section 6. 

\section{Preliminaries}

A static metric with cylindrical symmetry can be written in the coordinate
system $(x^{\mu})$ $\mu =0,1,2,3$ as
\begin{equation}
\label{4}
ds^2 =g_{00}dt^2 -g_{11}[(dx^1 )^2 +(dx^2 )^2 ]-g_{33}(dx^3 )^2
\end{equation}
where the $g_{00}$, $g_{11}$ and $g_{33}$ depend only on $\rho$
with $\rho =\sqrt{(x^1 )^2 +(x^2 )^2}$. 

Since we seek electromagnetic effects depending on $G$ in the first
approximation, we develop the
components $g_{\mu \nu}$ of the metric (\ref{4}) with respect to the
Minkowskian metric $\eta_{\mu \nu}$ as
\begin{equation}
\label{5}
g_{\mu \nu}=\eta_{\mu \nu}+h_{\mu \nu}
\end{equation}
where $h_{\mu \nu}$ are the linear terms in $G$. According to (\ref{4}),
 we see that
\begin{equation}
\label{6}
h_{11}=h_{22} \quad {\rm and} \quad h_{0i}=0
\end{equation}
The linearised Einstein equations for $h_{\mu \nu}$ take
the following form in which we must take into account the staticity and
the cylindrical symmetry
\begin{equation}
\label{6a}
\triangle  h_{\mu \nu}+\partial_{\mu}(\eta^{\sigma \rho}
\partial_{\sigma}h_{\nu \rho}-\frac{1}{2}\partial_{\nu}h)+\partial_{\nu}(
\eta^{\sigma \rho}\partial_{\sigma}h_{\mu \rho}-\frac{1}{2}\partial_{\mu}h)
=16\pi G(T_{\mu \nu}-\frac{1}{2}\eta_{\mu \nu}T)
\end{equation}
where $h=\eta^{\mu \nu}h_{\mu \nu}$, $\triangle$ being the Laplacian operator
which reduces here to $\triangle_{2}$ the Laplacian operator in two
dimensions. The energy-momentum tensor $T_{\mu \nu}$
which is the source of the gravitational field is the one calculated 
for a material distribution in the Minkowskian spacetime. It is assumed
to be regular.

The metric of an extended cosmic string is a particular case of metric 
(\ref{4}). In the limit case where its radius tends to zero, we get the
conical metric with a singular source.
In the linear approximation in $G$, the conical
metric can be written in the present coordinates \cite{vi3} as
\begin{equation}
\label{1}
ds^2 =(dx^0 )^2 -(1-8G\mu \ln \frac{\rho}{\rho_{*}})
[(dx^1 )^2 +(dx^2 )^2 ]-(dx^3 )^2
\end{equation}
where $\mu$ is the linear mass density of the cosmic string. We have 
introduced in (\ref{1}) an arbitrary length $\rho_{*}$ which can eventually
 represent the radius of the cosmic string.

\section{Equation of the electrostatic potential}

For electrostatics in a static metric, we use the electrostatic potential $V$
to which the vector potential $A_{\mu}$ reduces in this case
\begin{equation}
A_{0}=V \quad {\rm and} \quad A_{i}=0
\end{equation}
and it does not depend on the coordinate $x^{0}$. From the Maxwell equations
in the background metric (\ref{4}),
we obtain the equation of the electrostatic potential
\begin{equation}
\label{7}
\partial_{i}[\sqrt{-g}g^{ij}g^{00}\partial_{j}V]=
\frac{\sqrt{-g}}{\epsilon_{0}}j^{0}
\end{equation}
where $j^{0}$ is the charge density. For a point charge $q$ located at 
$x^{i}=x^{i}_{0}$, we have
\begin{equation}
\label{8}
j^{0}(x^{i})=\frac{q}{\sqrt{-g}}\delta^{(3)}(x^{i}-x^{i}_{0})
\end{equation}
The electrostatic potential $V$ has to be symmetric in the variables $x^{i}$
and $x_{0}^{i}$.

We can introduce development (\ref{5}) of the metric (\ref{4}) in
equation (\ref{7}). It is convenient to use the following notations
\begin{equation}
\label{9}
g^{00}=1+h^{00} \quad {\rm and} \quad \sqrt{-g}g^{ij}=-\delta^{ij}
+\gamma^{ij}
\end{equation}
where $h^{00}$ and $\gamma^{ij}$ are linear terms in $G$ given by
\begin{equation}
\nonumber \gamma^{ij}=h^{ij}-\frac{1}{2}h\delta^{ij} \quad {\rm and}
\quad h^{\mu \nu}=-\eta^{\mu \sigma}\eta^{\nu \rho}h_{\sigma \rho}
\end{equation}
It will be useful for determining the electrostatic potential to recast
$\gamma^{ij}$ in the form
\begin{equation}
\label{10}
\gamma^{ij}=\gamma_{\rm C}^{ij}+\gamma^{11}\delta^{ij} \quad {\rm with}
\quad \gamma_{\rm C}^{ij}=\delta^{i}_{3}\delta^{j}_{3}(\gamma^{33}
-\gamma^{11})
\end{equation}
which is possible taking into account (\ref{6}). We remark that metric 
(\ref{1}) corresponds to $h^{00}=0$ and 
$\gamma_{\rm C}^{ij}=\gamma_{\rm CS}^{ij}$ where $\gamma_{\rm CS}^{ij}$ is
defined by $\gamma_{\rm CS}^{33}=8G\mu \ln \rho /\rho_{*}$.

According to (\ref{10}), equation (\ref{7}) with source term (\ref{8})
can be now written  in the following manner
\begin{equation}
\label{11}
\partial_{i}[(-\delta^{ij}-\delta^{ij}(h^{00}-\gamma^{11})+
\gamma_{\rm C}^{ij})\partial_{j}V]=\frac{q}{\epsilon_{0}}
\delta^{(3)}(x^{i}-x^{i}_{0})
\end{equation}
In order to solve (\ref{11}), we write down $V$ as the sum of the Coulombian
potential and terms linear in $G$. We set
\begin{equation}
\label{12}
V(x^{i},x_{0}^{i})=\frac{q}{4\pi \epsilon_{0}r(x^{i},x_{0}^{i})}+
V_{\rm U}(x^{i},x_{0}^{i})+V_{\rm C}(x^{i},x_{0}^{i})
\end{equation}
where $r(x^{i},x_{0}^{i})$ is the Euclidean distance between the points
$x^{i}$ and $x^{i}_{0}$. If
$V_{\rm U}$ and $V_{\rm C}$ obey respectively the following equations
\begin{equation}
\label{13}
\triangle V_{\rm U}+\frac{q}{4\pi \epsilon_{0}}\partial_{i}
[(h^{00}-\gamma^{11})\delta^{ij}\partial_{j}\frac{1}{r}]=0
\end{equation}
\begin{equation}
\label{14}
\triangle V_{\rm C}-\frac{q}{4\pi \epsilon_{0}}\partial_{i}
[\gamma_{\rm C}^{ij}\partial_{j}\frac{1}{r}]=0
\end{equation}
then equation (\ref{11}) is satisfied.
We recall that we must find $V_{\rm U}$
and $V_{\rm C}$ symmetric in the variables $x^{i}$ and $x^{i}_{0}$.

For a straight cosmic string, the exact electrostatic potential is known
in another coordinate system in which the metric is conical. A local
expression is given for $G\mu<1/8$ \cite{li3} as the sum of the 
Coulombian potential and a regular part assuming that point $x$ is near point
$x_{0}$. At the linear approximation in $G$, a change of coordinate system
to obtain form (\ref{1}) of the metric yields simply
\begin{equation}
\label{14a}
V(x^{i},x_{0}^{i})=\frac{q}{4\pi \epsilon_{0}
r(x^{i},x_{0}^{i})}+V^{*}(x^{i},x_{0}^{i})
\end{equation}
valid only for a point $x^{i}$ near the point $x_{0}^{i}$ where the part
$V^{*}$ is regular at
$x^{i}=x_{0}^{i}$. It furnishes a solution $V_{\rm C}=V^{*}$ to equation
(\ref{14}) for $\gamma_{\rm C}^{ij}=\gamma_{\rm CS}^{ij}$. 
The self-interaction potential energy $W_{\rm CS}$ is then given by
\begin{equation}
\label{14b}
W_{\rm CS}(x_{0}^{A})=\frac{q}{2}V^{*}(x_{0}^{i},x_{0}^{i})
\end{equation} 
which does not depend on $x_{0}^{3}$. It has been explicitly 
calculated and it does not depend on the coordinate system 
at the linear approximation in $G$. We have
\begin{equation}
\label{3}
W_{\rm CS}(x_{0}^{A})=\frac{q^{2}G\mu}{16\epsilon_{0}\rho_{0}}
\end{equation} 
As expected, it is a function of $\rho_{0}$ with 
$\rho_{0}=\sqrt{(x_{0}^{1})^{2} +(x_{0}^{2})^{2}}$.

\section{Electrostatic self-force induced by $V_{\rm U}$}

We follow the method of Vilenkin \cite{vi1} for determining $V_{\rm U}$.
We seek the solution to equation (\ref{13}) in the form
\begin{eqnarray}
\label{15}
\nonumber & &V_{\rm U}(x^{i},x_{0}^{i})=-\frac{q}{8\pi \epsilon_{0}}
\left( h^{00}(x_{0}^{i})-\gamma^{11}(x_{0}^{i})\right)
\frac{1}{r(x^{i},x_{0}^{i})} \\
& &-\frac{q}{8\pi \epsilon_{0}}\left( h^{00}(x^{i})
-\gamma^{11}(x^{i})\right) \frac{1}{r(x^{i},x_{0}^{i})} \\
\nonumber & &+\psi (x^{i},x_{0}^{i})
\end{eqnarray}
Then, the quantity $\psi$ satisfies
\begin{equation}
\label{16}
\triangle \psi =\frac{q}{8\pi \epsilon_{0}r(x^{i},x_{0}^{i})}\triangle 
(h^{00}-\gamma^{11})
\end{equation}
The solution to equation (\ref{16}) is
\begin{equation}
\label{17}
\psi (x^{i},x_{0}^{i})=-\frac{q}{32\pi^{2}\epsilon_{0}}\int 
\frac{\triangle (h^{00}-\gamma^{11})(b^{i})}
{r(b^{i},x_{0}^{i})r(b^{i},x^{i})}d^{3}b
\end{equation}
which is symmetric in $x^{i}$ and $x_{0}^{i}$.

From (\ref{17}), we immediately obtain the self-interaction potential energy
\begin{equation}
W_{\rm U}(x_{0}^{i})=\frac{q}{2}\psi (x_{0}^{i},x_{0}^{i})
\end{equation}
 which is regular in $x^{i}=x_{0}^{i}$. It is
given by the integral
\begin{equation}
\label{18}
W_{\rm U}(x_{0}^{i})=-\frac{q^{2}}{64\pi^{2}\epsilon_{0}}\int
\frac{\triangle (h^{00}-\gamma^{11})(b^{i})}{r^{2}(b^{i},x_{0}^{i})}d^{3}b
\end{equation}
which is well defined.

In the case of the cylindrical symmetry, $\triangle$ acting on 
$h^{00}-\gamma^{11}$
is in fact $\triangle_{2}$ the Laplacian operator in two dimensions.
 Moreover, we can perform in (\ref{18}) the integration
on the variable $b^{3}$ and we obtain
\begin{equation}
\label{19}
W_{\rm U}(x_{0}^{A})=-\frac{q^{2}}{64\pi \epsilon_{0}}\int
\frac{\triangle_{2}(h^{00}-\gamma^{11})(b^{A})}
{\rho (x_{0}^{A},b^{A})}d^{2}b
\end{equation}
where $\rho (x_{0}^{A},b^{A})=\sqrt{(x_{0}^{1}-b^{1})^2 +(x_{0}^{2}-b^{2})^2}$.
As a result of the integration on $b^{3}$, $W_{\rm U}$ does not depend on
$x^{3}_{0}$. Futhermore, it is easy to see that $W_{\rm U}$ is a function 
of $\rho_{0}$.

In general relativity, we may use the linearised Einstein equations (\ref{6a}) 
having as source term the energy-momentum tensor
$T_{\mu \nu}$ to calculate \\
$\triangle_{2}(h^{00}-\gamma^{11})$. After some manipulations, we find
\begin{equation}
\label{20}
W_{\rm U}(x_{0}^{A})=\frac{q^{2}G}{8\epsilon_{0}}\int
\frac{(T^{0}_{0}-T^{3}_{3})(b^{A})}{\rho (x_{0}^{A},b^{A})}d^{2}b
\end{equation}

\section{Electrostatic self-force induced by $V_{\rm C}$}

We now turn our attention to determining $V_{\rm C}$.
We can solve equation (\ref{14}) if we know the solution 
$G(x^{i},x_{0}^{i};b^{A})$ to the following equation
\begin{equation}
\label{21}
\triangle G-\frac{q}{4\pi \epsilon_{0}}
\partial_{i}\left[ \delta_{3}^{i}\delta_{3}^{j}\delta^{(2)}(x^{A}-b^{A})
\partial_{j}\frac{1}{r(x^{i},x_{0}^{i})}\right] =0
\end{equation}
the desired solution will be written in an integral form
\begin{equation}
\label{22}
V_{\rm C}(x^{i},x_{0}^{i})=\int \gamma_{\rm C}^{33}(b^{A})
G(x^{i},x_{0}^{i};b^{A})d^{2}b
\end{equation}
We have already noticed that $V^{*}$, appearing in (\ref{14a}) for
a straight cosmic string, satisfies (\ref{14}) which in this particular
case takes the form
\begin{equation}
\label{23}
\triangle V^{*}-\frac{2G\mu q}{\pi \epsilon_{0}}
\partial_{i}\left[ \delta^{i}_{3}\delta^{j}_{3}\ln \frac{\rho}{\rho_{*}}
\partial_{j}\frac{1}{r(x^{i},x_{0}^{i})}\right] =0
\end{equation}
As mentioned this solution is known for $x^{i}$ near $x_{0}^{i}$.
A change of origin of the coordinates $(x^{A})$ by $b^{A}$ in 
equation (\ref{23}) yields the equation
\begin{equation}
\label{24}
\triangle V^{*}_{b}-\frac{2G\mu q}{\pi \epsilon_{0}}
\partial_{i}\left[ \delta^{i}_{3}\delta^{j}_{3}\ln \frac{\rho (x^{A},b^{A})}
{\rho_{*}}\partial_{j}\frac{1}{r(x^{i},x_{0}^{i})}\right] =0
\end{equation}
whose solution $V^{*}_{b}$ is consequently given by
\begin{equation}
\label{24a}
V^{*}_{b}(x^{i},x_{0}^{i};b^{A})=
V^{*}(x^{A}-b^{A},x^{3};x^{A}_{0}-b^{A},x_{0}^{3})
\end{equation}
But $\triangle_{2}\ln \rho /\rho_{*}=2\pi\delta^{(2)}$, therefore
\begin{equation}
\label{25}
G(x^{i},x_{0}^{i};b^{A})=\frac{1}{16\pi G\mu}\triangle_{2}
V^{*}_{b}(x^{i},x_{0}^{i};b^{A})
\end{equation}
where $\triangle_{2}$ acts on the variables $b^{A}$. According to (\ref{22}),
we are now in a position 
to give the integral expression of the desired solution
\begin{equation}
\label{26}
V_{\rm C}(x^{i},x_{0}^{i})=\frac{1}{16\pi G\mu}\int
\triangle_{2}\gamma^{33}_{\rm C}(b^{A})V^{*}_{b}(x^{i},x_{0}^{i};b^{A})d^{2}b
\end{equation}
for a point $x^{i}$ near point $x_{0}^{i}$, in which we have used a property of
the convolution product so that
$\triangle_{2}$ applies now on $\gamma_{\rm C}^{33}$.
Expression (\ref{26})  is regular at $x^{i}=x_{0}^{i}$.

From (\ref{26}), we deduce immediately the self-interaction potential energy
\begin{equation}
\label{28}
W_{\rm C}(x_{0}^{A})=\frac{q}{32\pi G\mu}\int
\triangle_{2}\gamma^{33}_{\rm C}(b^{A})
V^{*}_{b}(x_{0}^{i},x_{0}^{i};b^{A})d^{2}b
\end{equation}
By considering the self-interaction potential energy 
derived from (\ref{3}) for a straight 
cosmic string located at $x^{A}=b^{A}$ 
\begin{equation}
\label{28a}
\frac{1}{2}qV_{b}^{*}(x^{i}_{0},x_{0}^{i};b^{A})=
\frac{q^{2}G\mu}{16\epsilon_{0}\rho (x_{0}^{A},b^{A})}
\end{equation}
we get from (\ref{28}) an integral expression
\begin{equation}
\label{29}
W_{\rm C}(x_{0}^{A})=\frac{q^{2}}{256\pi \epsilon_{0}}\int
\frac{\triangle_{2}\gamma^{33}_{\rm C}(b^{A})}
{\rho (x_{0}^{A},b^{A})}d^{2}b
\end{equation}

In general relativity we may use the linearised Einstein equations (\ref{6a})
to calculate $\triangle_{2}\gamma_{\rm C}^{33}$ ;
we obtain after some manipulations
\begin{equation}
\label{30}
W_{\rm C}(x_{0}^{A})=\frac{q^{2}G}{16\epsilon_{0}}\int
\frac{(T^{3}_{3}-T^{1}_{1}-T^{2}_{2})(b^{A})}
{\rho (x_{0}^{A},b^{A})}d^{2}b
\end{equation}

\section{Conclusion} 

For a static spacetime with cylindrical symmetry, we have determined in
the linear approximation in $G$ the self-interaction potential energy
$W_{\rm U}+W_{\rm C}$ given respectively by (\ref{19}) and (\ref{29}) in terms
of the components of the metric. We can apply these formulas for all
gravitational metric theories.

Taking into account the linearised Einstein equations, we have expressed these
potentials with the aid of the components of the energy-momentum tensor
which is the source of the gravitational field. We have found the following
self-interaction potential energy
\begin{equation}
\label{31}
W(x_{0}^{A})=\frac{q^{2}G}{8\epsilon_{0}}\left[ \int
\frac{T^{0}_{0}(b^{A})}{\rho (x_{0}^{A},b^{A})}d^{2}b
-\frac{1}{2}\int
\frac{(T^{3}_{3}+T^{1}_{1}+T^{2}_{2})(b^{A})}
{\rho (x_{0}^{A},b^{A})}d^{2}b\right]
\end{equation}
which depends only on $\rho_{0}$. 
For an energy-momentum tensor vanishing at infinity, we point out that 
the integrals in expression (\ref{31}) are well defined. We emphasize that
formula (\ref{31}) can be now evaluated in arbitrary 
Minkowskian-like coordinates in the linear approximation in $G$.

Formula (\ref{31}) is general for regular energy-momentum tensors. 
It can be applied for different types of cosmic string.
This formula obviously verifies the limit case in which the radius
of an extented cosmic string tends to zero since 
\begin{equation}
\nonumber T_{0}^{0}\rightarrow \mu \delta^{(2)} \quad 
T_{3}^{3}\rightarrow \mu \delta^{(2)}
\quad T_{1}^{1}\rightarrow 0 \quad T_{2}^{2}\rightarrow 0
\end{equation}
giving expression (\ref{3}) for an infinitely thin cosmic string.

\end{document}